\documentclass[pra,superscriptaddress,twocolumn,amsmath,amssymb]{revtex4-1}	
\usepackage{graphicx}
\usepackage{dcolumn} 
\usepackage{bm}
\usepackage{epsfig}
\usepackage{braket}
\usepackage{hyperref}
\hypersetup{
    colorlinks=true,        
    linkcolor=blue,          
    citecolor=blue,        
    filecolor=magenta,      
    urlcolor=blue           
}

\def\laco*{LaCoO$_3$}
\def\ehs*{$\epsilon_{\text{HS}}$}
\def\eis*{$\epsilon_{\text{IS}}$}
\def\els*{$\epsilon_{\text{LS}}$}

\newcommand{\citenamefont}{}
\newcommand{\bibnamefont}{}
\def\s*{$|s\rangle$}
\def\t*{$|t_\alpha\rangle$}
\def\se*{$s$}
\def\te*{$t_{\alpha}$}
\def\jo*{$j_{1/2}$}
\def\jt*{$j_{3/2}$}

\begin{document}
\title{Competing phases in a model of Pr-based cobaltites}

\author{A. Sotnikov}
\email{sotnikov@ifp.tuwien.ac.at}
\affiliation{Institute of Solid State Physics,
TU Wien, Wiedner Hauptstr. 8, 1020 Vienna, Austria}
\author{J. Kune\v{s}}
\affiliation{Institute of Solid State Physics,
TU Wien, Wiedner Hauptstr. 8, 1020 Vienna, Austria}
\affiliation{Institute of Physics, ASCR, Na Slovance 2,
182 21 Praha 8, Czech Republic}

\date{\today}

\begin{abstract}
Motivated by the physics of Pr-based cobaltites, we study the effect of the external magnetic field in the hole-doped two-band Hubbard model close to instabilities toward the excitonic condensation and ferromagnetic ordering. Using the dynamical mean-field theory we observe a field-driven suppression of the excitonic condensate. The onset of a magnetically ordered phase at the fixed chemical potential is accompanied by a sizable change of the electron density. This leads us to predict that Pr$^{3+}$ abundance increases on the high-field side of the transition.

\end{abstract}
\maketitle

\section{Introduction}
The proximity of the Co$^{3+}$ ionic state in LaCoO$_3$ and related compounds to the spin-state transition gives rise to a number
of unusual physical properties, which have continued to attract attention for over 50 years. The small energy gap separating
spinful excitations from the singlet [low-spin (LS)] ground state of Co$^{3+}$ ion leads to  
a broad crossover of LaCoO$_3$  from a nonmagnetic insulator to a paramagnetic Curie-Weiss insulator 
(and, eventually, metal) with increasing temperature. The thermal population of the excited atomic multiplets
of Co leaves numerous signatures in spectroscopy, e.g., in valence photoemission~\cite{saitoh97} or Co $L$-edge
x-ray absorption~\cite{haverkort06}, or results in anomalous expansion of Co-O bonds. Generally, it is accepted
that at elevated temperatures the atomic states of Co, LS, high spin (HS), or intermediate spin (IS),
acquire fractional populations. In materials where the spin-state transition is complete,
it is usually of the first-order type accompanied by an abrupt volume change~\cite{guetlich2004spin}.
In some cases, the metal-insulator transition takes place simultaneously~\cite{yoo05,gavriliuk08,kunes08}.

Materials from the  (Pr$_{1-y}$Ln$_{y}$)$_{x}$Ca$_{1-x}$CoO$_3$ (PCCO) family, where Ln is a trivalent ion (Ln$=$Y, Sm, Gd, {etc.}), 
undergo a transition from a high-temperature Curie-Weiss metal to a low-temperature insulator without Co local moments signature for $x\ge0.5$.
Unlike the broad crossover of LaCoO$_3$, a sharp peak in the specific heat clearly points to the collective nature of
the transition in PCCO~\cite{tsubouchi02,tsubouchi04,hejtmanek13}.
The high-$T$ phase of PCCO corresponds to a heavily hole-doped cobaltite, with Co formal valence $3+x$.
The Pr$^{3+}$ to Pr$^{4+}$ valence transition observed simultaneously with other changes of physical characteristics puts the low-$T$ phase
much closer to the Co$^{3+}$ formal valence. The PCCO transitions lack the anomalous change of the Co-O bond length~\cite{fujita04}
and some x-ray absorption signature of the spin-state crossover~\cite{herrero12} observed in LaCoO$_3$. 
Most importantly, the low-$T$ phase of PCCO breaks the time-reversal symmetry as reflected in the splitting
of the Kramers ground state of Pr$^{4+}$ ions~\cite{hejtmanek10,hejtmanek13,knizek13}. These observations indicate that PCCO does not undergo
a crossover between different physical regimes as does LaCoO$_3$, but a phase transition with spontaneous 
symmetry breaking.

Kune\v{s} and Augustinsk\'y~\cite{kunes14b} proposed that PCCO undergoes a condensation of spinful excitons,
similar to excitonic magnetism~\cite{keldysh65,Halperin1968,balents00,kha13} reported in $d^4$ ruthenates~\cite{jain17}.
In a material with a singlet {\it atomic} ground state and a small excitation energy of the lowest spinful multiplet, 
interatomic exchange processes give rise to a {\it global} ground state with spontaneously broken symmetry --- excitonic condensate (EC).
In the condensate, the low-energy atomic states form a coherent superposition, which distinguishes it from
a normal state with fractional atomic state populations due to thermal excitations.
Dynamical mean-field theory (DMFT) calculations for a simplified two-orbital model and 
material-specific density-functional (LDA+$U$) calculations~\cite{kunes14b} 
captured the essentials of the PCCO physics, including metal-insulator transition, vanishing Curie-Weiss response of 
Co ions, connection to the Pr valence transition, and the splitting of the Pr$^{4+}$ ground-state doublet. 

The ultimate proof of the EC scenario for the PCCO may come from the two-particle
excitation spectra. Yamaguchi {\it et al.}~\cite{yamaguchi17} 
observed the excitonic instability and computed the spin excitation spectra in a realistic five-orbital model 
using the weak-coupling Hartree-Fock and random phase approximations.
Nasu {\it et al.}~\cite{nasu16} obtained the mean-field phase diagram 
and the excitation spectrum in the linear spin-wave approximation for
the strong-coupling limit of the two-orbital Hubbard model. The experimental investigations are highly desirable.

Recently, the high-field experiments on PCCO~\cite{naito14,ikeda16b} revealed that the low-temperature phase is suppressed by magnetic field.
In Refs.~\cite{sotnikov16sr,tatsuno16} the effect of magnetic field was studied in the two-orbital model in the vicinity of 
excitonic instability and it was shown that a large enough field induces the excitonic condensation (that was consistent with the experiment~\cite{ikeda16a} in LaCoO$_3$).
However, this observation appears to contradict the excitonic scenario for PCCO. The purpose of this paper is to show that it is not the case.

\section{Model and Computational Method} 
In order to focus on the essential physics as well as to reduce the computational effort, we use a minimal two-band Hubbard model (2BHM) on square lattice. 
Numerous studies of 2BHM revealed excitonic instability for suitable parameters at half-filling, $n=2$~\cite{bronold06,brydon09,kaneko12,kunes15,hoshino16,nasu16,kaneko16}.
Doping suppresses the excitonic condensate~\cite{balents00,kunes14c,kunes16prl} and turns it into a paramagnetic metal at some critical charge density. 
Pushing the doping further, one eventually arrives at a ferromagnetic (FM) phase~\footnote{This phase is adiabatically connected to the double-exchange regime at filling $n=1+\delta$.}.
By moderate changes of the model parameters from Ref.~\cite{kunes14c} (increasing the interaction strength~$U$ and the bands asymmetry),
the excitonic and the ferromagnetic regions can be expanded such that they come into contact. 
This is the parameter regime we address. 

The Pr ions play an important role in the physics of PCCO providing a charge reservoir for the active Co-derived bands. 
The delicate balance between the Pr$^{3+}$ and Pr$^{4+}$ valence states fixes the chemical potential of the Co $d$ bands.

The Hamiltonian of the two-orbital model in an external magnetic field~$B$ reads
\begin{eqnarray}
 {\cal H} &=& \sum_{\alpha\beta}t_{\alpha\beta}\sum_{\braket{ij}\sigma} (c^\dag_{i\alpha\sigma}c^{\phantom{\dag}}_{j\beta\sigma} + \text{H.c.})
 \nonumber\\
 && 
 + \sum_{i}{\cal H}^{(i)}_{\text{int}}
 + \sum_{i\alpha\sigma}(\sigma B+\epsilon_\alpha)n_{i\alpha\sigma},\label{H}
\end{eqnarray}
where $c^\dag_{i\alpha\sigma}$ ($c_{i\alpha\sigma}$) are fermionic operators creating (annihilating) an electron with
the spin $z$ projection $\sigma=\pm 1$ on the orbital $\alpha=\{a,b\}$ at the lattice site $i$. 
The operator $n_{i\alpha\sigma}=c^\dag_{i\alpha\sigma}c^{\phantom\dagger}_{i\alpha\sigma}$ denotes the corresponding density.
The external magnetic field $B$ is assumed to act on the spin only. The zero-field on-site energies, $\epsilon_a=\mu$ and $\epsilon_b=\mu-\Delta$,
contain the chemical potential $\mu$ and the crystal-field splitting $\Delta$. The 
$2\times2$ symmetric matrix $t_{\alpha\beta}$ contains of the nearest-neighbor hopping amplitudes on a square lattice. 
 The local interaction part~${\cal H}^{(i)}_{\text{int}}$ is chosen to have only density-density contributions,
\begin{eqnarray}
 {\cal H}^{(i)}_{\text{int}} &=& U\sum_{\alpha} n_{i\alpha\uparrow}n_{i\alpha\downarrow} + (U-2J)\sum_{\sigma} n_{ia\sigma}n_{ib-\sigma}
 \nonumber\\
 &&
 +(U-3J)\sum_{\sigma} n_{ia\sigma}n_{ib\sigma}.
\end{eqnarray}

We use DMFT~\cite{Geo1996RMP} with the continuous-time quantum Monte Carlo hybridization-expansion (CT-HYB) impurity solver \cite{Wer2006PRL,Gul2011RMP} and off-diagonal hybridization, which allows description of the ordered EC state  \cite{kunes14a,kunes14c}.

The excitonic phase under study is characterized by the uniform order parameter 
$\boldsymbol{\phi}=\sum_{\sigma\sigma'}\boldsymbol{\tau}_{\sigma\sigma'}\langle c^\dag_{ia\sigma}c^{\phantom\dagger}_{ib\sigma'}\rangle$, 
where $\boldsymbol{\tau}=(\tau^x,\tau^y, \tau^z)$ are the Pauli matrices. 
With the present restriction to density-density interactions, $\boldsymbol{\phi}$ is confined in the $xy$ plane, 
thus we compute its helical components $\phi^{+}=\braket{c^\dag_{ia\uparrow}c^{\phantom\dagger}_{ib\downarrow}}$ 
and $\phi^{-}=\braket{c^\dag_{ia\downarrow}c^{\phantom\dagger}_{ib\uparrow}}$. In addition, we follow
the magnetization $m=\sum_{\sigma\alpha}\sigma \braket{n_{i\alpha\sigma}}$ and the electron density $n=\sum_{\sigma\alpha} \braket{n_{i\alpha\sigma}}$.
We point out that the orientation of the magnetic field along the $z$ axis is consistent with the $xy$-plane orientation of the EC order parameter. 
In a rotationally symmetric setting, the order parameter  $\boldsymbol{\phi}$ orients itself perpendicular to an external field, similarly to the behavior 
of antiferromagnetic polarization (i.e., forming the canted configuration) in magnetic field.

\section{Results and Discussion}
\begin{figure}
\includegraphics[width=\linewidth]{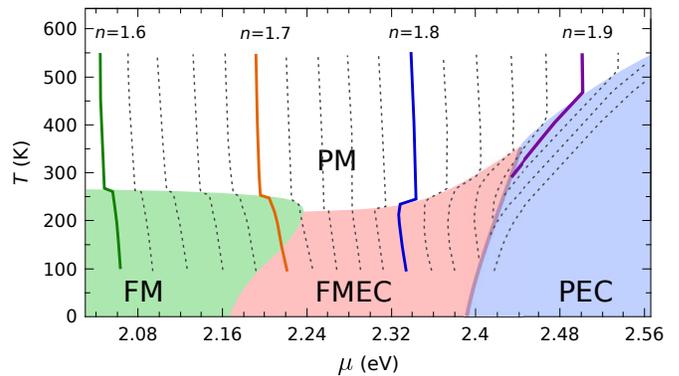}
    \caption{\label{fig1}
    Phase diagram of the model, including polar EC (PEC), ferromagnetic EC (FMEC), ferromagnetic (FM), and paramagnetic (PM) phases at zero magnetic field.
    The dependence $n(\mu,T)$ is shown by means of contour lines.
    The phase boundaries below $97$~K are obtained from extrapolation. 
    The Hubbard parameters (given in units of eV) are $\Delta=3.2$, $U=5$, $J=1$, $t_{aa}=0.4284$, $t_{bb}=-0.1466$, and $t_{ab}=0.02$.}
\end{figure}

\begin{figure}
\includegraphics[width=\linewidth]{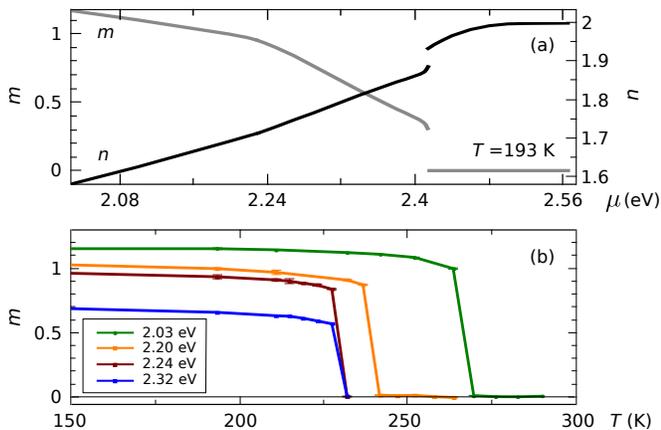}
    \caption{\label{figMdep} Dependencies of the electron density~$n$ and the magnetization~$m$ on the chemical potential~$\mu$ at $T=193$~K (a), 
    and dependencies of the magnetization at different $\mu$ on the temperature $T$ (b).
    Other Hubbard parameters are taken the same as in Fig.~\ref{fig1}.}
\end{figure}

First, we discuss the phase diagram in the absence of an external field shown in Fig.~\ref{fig1}. 
Besides the normal paramagnetic (PM) phase, we observe two distinct EC phases, the polar (PEC) and ferromagnetic (FMEC) condensates~\cite{ho98,ohmi98}, and the conventional ferromagnetic phase.
We observe continuous phase transitions at FM/FMEC and PEC/PM boundaries. 
The PEC/FMEC transition is of the first order due to charge separation (see also Ref.~\cite{kunes14c} for details). 
This is reflected in the collapse of constant-density contours onto the single line in Fig.~\ref{fig1}. 
The jump in the charge density~$n$ and magnetization $m$ at the transition is shown in Fig.~\ref{figMdep}(a).
A stepwise change of magnetization is observed also at the PM/FM and PM/FMEC transitions [see Fig.~\ref{figMdep}(b)].
The lines of constant density in Fig.~\ref{fig1} indicate that at a constant chemical potential the electron density increases with $T\to0$ on the lightly doped (right) side, while the behavior is opposite on the heavily doped (left) side of the diagram. 
This originates from different physical requirements for optimal dopings: the PEC phase is most stable at stoichiometric filling ($n=2$), while the FM phase requires a large number of mobile carriers necessary for stabilization of the double-exchange mechanism ($n\lesssim1.5$). Therefore, in the intermediate (FMEC) regime we observe the corresponding change in the temperature behavior of the constant-density lines.

Discussing the symmetry aspects of the phase diagram in Fig.~\ref{fig1}, it is instructive to assume that the model
has full SU(2) spin-rotational symmetry. Breaking of continuous symmetries has observable consequences in Nambu-Goldstone (NG) modes between the various phases. 
Following the analysis of Refs.~\cite{nasu16,bascones02,watanabe12}, the PEC phase has a residual U(1) symmetry 
and breaks two generators with vanishing expectation value of their commutator.
The FMEC phase has no residual continuous symmetry and thus breaks all three generators with one nonzero commutator (more precisely, the matrix of commutators has a rank 2).
The FM phase has U(1) residual symmetry and breaks two generators with finite expectation value of their commutator.
Therefore, there are two NG modes with linear dispersion in the PEC phase, one linear and one quadratic NG mode in the FMEC phase, and one
quadratic NG mode in the FM phase~\cite{watanabe12}.

\begin{figure}
\includegraphics[width=\linewidth]{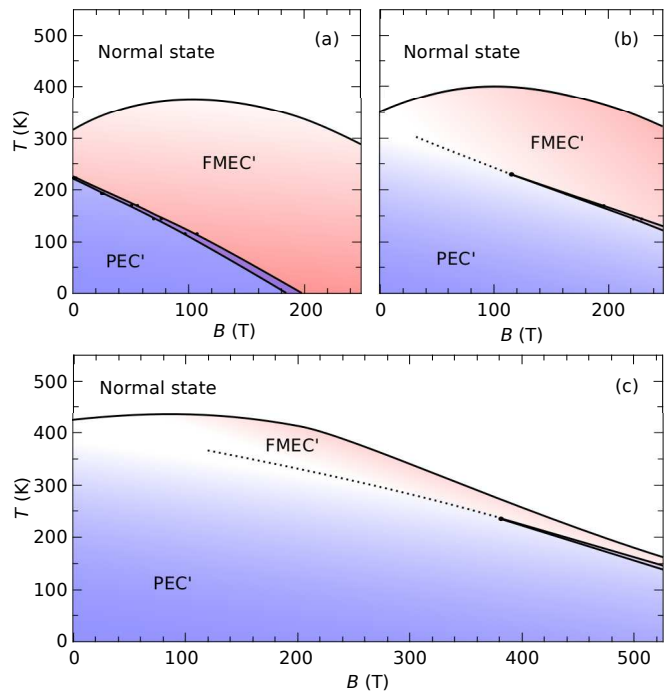}
    \caption{\label{figBTpd} 
    $B$-$T$ phase diagrams for three chemical potential values, $\mu=2.42$~eV (a), $\mu=2.44$~eV (b), and $\mu=2.48$~eV (c). 
    Other Hubbard parameters are taken the same as in Fig.~\ref{fig1}.}
\end{figure}

\begin{figure}
\includegraphics[width=\linewidth]{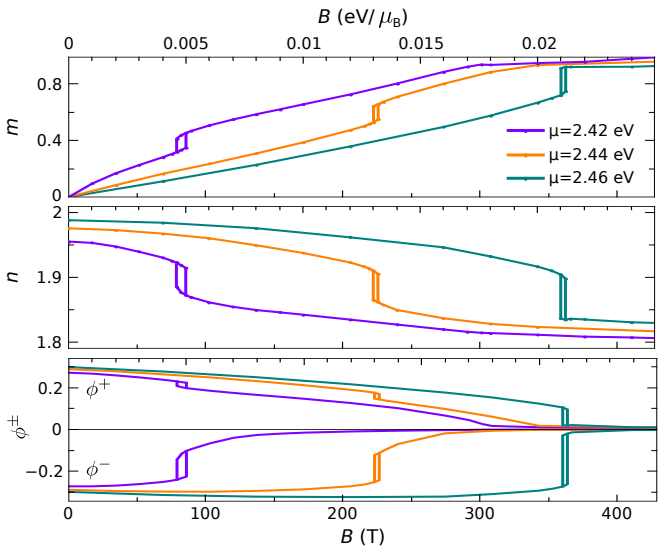}
    \caption{\label{figBfield}
    Dependencies of magnetization~$m$, electron density~$n$, and EC order parameters~$\phi^{\pm}$ at constant temperature ($T=145$~K) and different chemical potential values ($\mu=2.42,~2.44,~2.46$~eV). Other Hubbard parameters are taken the same as in Fig.~\ref{fig1}.}
\end{figure}

\begin{figure}
\includegraphics[width=\linewidth]{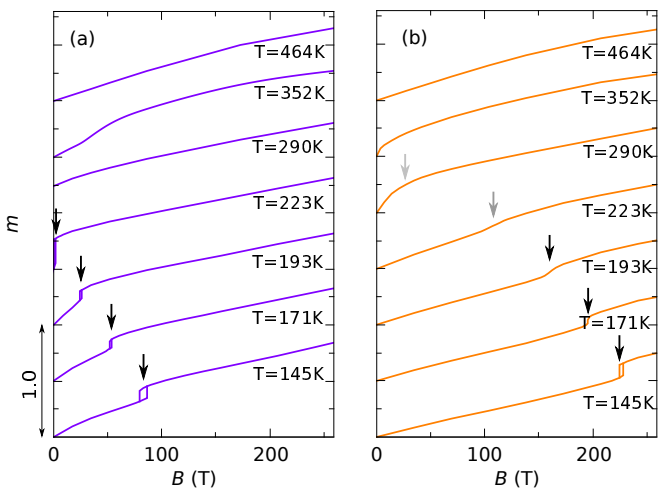}
    \caption{\label{figTdep}
    Dependencies of magnetization at different temperatures for two chosen chemical potential values: $\mu=2.42$\,eV (a) and $\mu=2.44$\,eV (b). Other Hubbard parameters are taken the same as in Fig.~\ref{fig1}.}
\end{figure}

Next, we discuss the effect of magnetic field~$B$. The calculations were performed for several chemical potentials~$\mu$
close to the boundary of the PEC phase. 
At low temperatures we observe a first-order transition from the PEC$^\prime$ to FMEC$^\prime$ phase (not distinguished by symmetry at $B\neq0$)
with increasing field $B$ followed by a continuous transition to the normal state (see Fig.~\ref{figBTpd}). 
The latter transition is distinguished by symmetry at any $B$.
The first-order transition is accompanied by a stepwise increase of $m$ and drop of the electron density $n$,
as well as a small hysteresis (see Fig.~\ref{figBfield}).
The change from FMEC' to normal phase leaves only a moderate kink in the $m(B)$ and $n(B)$ dependencies.
At higher temperatures the PEC'/FMEC' transition turns into a crossover as shown in Fig.~\ref{figTdep}. Similarly to the experiment~\cite{ikeda16b}, 
we observe a decrease of critical field with temperature, $dB_c/dT<0$. 

Finally, we briefly comment on the shape of the magnetization curves. The concave $m(B)$ dependence at low $B$,  
particularly clear when $B_c$ is small, is in contrast to the behavior of an isolated atom with the LS ground state and 
thermally populated HS states, which leads to convex $m(B)$ dependence (becoming almost linear at temperatures comparable
to the HS excitation energy). While other explanations, e.g., surface or impurity magnetism, are possible, we point out
that similar nonlinearities can also be noticed in the experimental data \cite{ikeda16b}.

The simplified model has necessarily limitations in describing real PCCO. In particular, the extent or presence of FMEC is exaggerated. Using a more realistic (and computationally
demanding) form of the interaction Hamiltonian including the spin-flip and pair-hopping terms is likely to suppress the extent of the FMEC phase, possibly placing the first-order
transition directly between the PEC and FM phases. The lattice response present in the real material is likely to enhance the hysteretic behavior.
However, two observations are general. First, an external magnetic field suppresses the EC phase with the consequences such as the onset
of metallic conduction. Second, the suppression of EC in a system with fixed chemical potential leads to substantial change of the electron density.
Therefore, we predict that the field-induced transition in PCCO is accompanied by Pr$^{4+}$ to Pr$^{3+}$ valence transition. Although similar
behavior is observed in the temperature-driven transition, this prediction is nontrivial. This is because a magnetic field acting on the Pr ions subject to
a fixed ligand field leads to an opposite effect, i.e., it favors the spinful Pr$^{4+}$ state rather than Pr$^{3+}$ with the singlet ground state.

\section{Conclusions}
We have studied the suppression of the excitonic condensate by magnetic field in the two-band Hubbard model.
The observed behavior qualitatively agrees with the experiments on PCCO.
We have shown that the field-induced transition at a fixed chemical potential is accompanied by a substantial change of the electron density.
Therefore, we predict that the field-driven suppression of the EC state results
in the Pr$^{4+}$ to Pr$^{3+}$ valence crossover. 

\begin{acknowledgments}
The authors thank A. Hariki and P. Nov\'{a}k for fruitful discussions.
This work has received funding from the European Research Council (ERC) under the European Union's Horizon 2020 research
and innovation program (Grant Agreement No. 646807-EXMAG).
Access to computing and storage facilities owned by parties and projects contributing to the National Grid Infrastructure 
MetaCentrum,  provided under the  program ``Projects  of Large Infrastructure for Research, Development, and 
Innovations'' (LM2010005), and the Vienna Scientific Cluster (VSC) is greatly appreciated.
\end{acknowledgments}

\bibliography{FMECfield}

\end{document}